\documentclass[aps,prl,twocolumn,superscriptaddress,showpacs,amsmath,amssymb]{revtex4-2}
\usepackage{graphicx} 
\usepackage{epstopdf} 
\usepackage{dcolumn} 
\usepackage{xcolor} 
\usepackage{amsmath,amssymb} 
\usepackage{hyperref} 
\usepackage[utf8]{inputenc} 
\usepackage[english]{babel} 
\usepackage[displaymath, mathlines]{lineno} 
\usepackage{blindtext} 
\usepackage{ifthen} 
\usepackage{orcidlink} 

\usepackage{subfig}
\usepackage{physics}
\usepackage{upgreek}
\usepackage{cleveref}
\crefname{figure}{Fig.}{Figs.}
\Crefname{figure}{Figure}{Figures}
\crefname{table}{Table}{Tables}
\crefname{equation}{Eq.}{Eqs.}
\usepackage[justification=raggedright,singlelinecheck=false]{caption}
\DeclareCaptionLabelFormat{myfig}{FIG. #2}
\DeclareCaptionLabelFormat{mytab}{TABLE #2}
\usepackage[normalem]{ulem} 

\graphicspath{{figures/}} 

\newboolean{articletitles}
\setboolean{articletitles}{true} 

\RequirePackage{xspace}

\def\belletwo {Belle~II\xspace}

\def\epem       {\ensuremath{e^+e^-}\xspace}

\def\mumu       {\ensuremath{\mu^+\mu^-}\xspace}

\def\tautau     {\ensuremath{\tau^+\tau^-}\xspace}

\def\g     {\ensuremath{\gamma}\xspace}

\def\qqbar {\ensuremath{q\overline q}\xspace}

\def\cbar  {\ensuremath{\overline c}\xspace}

\def\piz   {\ensuremath{\pi^0}\xspace}

\def\pip   {\ensuremath{\pi^+}\xspace}

\def\pipi  {\ensuremath{\pi^+\pi^-}\xspace}

\def\Kbar  {\kern 0.2em\overline{\kern -0.2em K}{}\xspace}

\def\Kz    {\ensuremath{K^0}\xspace}
\def\Kzb   {\ensuremath{\Kbar^0}\xspace}
\def\KzKzb {\ensuremath{\Kz \kern -0.16em \Kzb}\xspace}
\def\Kp    {\ensuremath{K^+}\xspace}
\def\Km    {\ensuremath{K^-}\xspace}

\def\KpKm  {\ensuremath{\Kp \kern -0.16em \Km}\xspace}
\def\KS    {\ensuremath{K^0_{\scriptscriptstyle S}}\xspace} 
 
\def\Kstarz  {\ensuremath{K^{*0}}\xspace}

\def\Dbar    {\kern 0.2em\overline{\kern -0.2em D}{}\xspace}

\def\Dz      {\ensuremath{D^0}\xspace}
\def\Dzb     {\ensuremath{\Dbar^0}\xspace}
\def\DzDzb   {\ensuremath{\Dz {\kern -0.16em \Dzb}}\xspace}
\def\Dp      {\ensuremath{D^+}\xspace}
\def\Dm      {\ensuremath{D^-}\xspace}

\def\DpDm    {\ensuremath{\Dp {\kern -0.16em \Dm}}\xspace}

\def\Bbar    {\kern 0.18em\overline{\kern -0.18em B}{}\xspace}

\def\BB      {\ensuremath{B\Bbar}\xspace} 
\def\Bz      {\ensuremath{B^0}\xspace}
\def\Bzb     {\ensuremath{\Bbar^0}\xspace}
\def\BzBzb   {\ensuremath{\Bz {\kern -0.16em \Bzb}}\xspace}
\def\Bu      {\ensuremath{B^+}\xspace}
\def\Bub     {\ensuremath{B^-}\xspace}
\def\Bp      {\ensuremath{\Bu}\xspace}

\def\BpBm    {\ensuremath{\Bu {\kern -0.16em \Bub}}\xspace}

\def\FourS {\ensuremath{\Upsilon(4S)}\xspace}

\mathchardef\Deltares="7101

\def\Deltabar{\kern 0.25em\overline{\kern -0.25em \Deltares}{}\xspace}
\def\Lbar{\kern 0.2em\overline{\kern -0.2em\Lambda\kern 0.05em}\kern-0.05em{}\xspace}
\def\Sigbar{\kern 0.2em\overline{\kern -0.2em \Sigma}{}\xspace}
\def\Xibar{\kern 0.2em\overline{\kern -0.2em \Xi}{}\xspace}
\def\Obar{\kern 0.2em\overline{\kern -0.2em \Omega}{}\xspace}
\def\Nbar{\kern 0.2em\overline{\kern -0.2em N}{}\xspace}
\def\Xb{\kern 0.2em\overline{\kern -0.2em X}{}\xspace}

\newcommand{\tev}{\ensuremath{\mathrm{~Te\kern -0.1em V}}\xspace}
\newcommand{\gev}{\ensuremath{\mathrm{~Ge\kern -0.1em V}}\xspace}
\newcommand{\mev}{\ensuremath{\mathrm{~Me\kern -0.1em V}}\xspace}
\newcommand{\kev}{\ensuremath{\mathrm{~ke\kern -0.1em V}}\xspace}
\renewcommand{\ev}{\ensuremath{\mathrm{~e\kern -0.1em V}}\xspace}
\newcommand{\gevc}{\ensuremath{{\mathrm{~Ge\kern -0.1em V\!/}c}}\xspace}
\newcommand{\mevc}{\ensuremath{{\mathrm{~Me\kern -0.1em V\!/}c}}\xspace}
\newcommand{\gevcc}{\ensuremath{{\mathrm{~Ge\kern -0.1em V\!/}c^2}}\xspace}
\newcommand{\mevcc}{\ensuremath{{\mathrm{~Me\kern -0.1em V\!/}c^2}}\xspace}

\def\cm   {\ensuremath{{\rm ~cm}}\xspace}

\def\mum  {\ensuremath{{~\upmu\rm m}}\xspace}

\def\invfb   {\ensuremath{\mbox{~fb}^{-1}}\xspace}

\def\mus  {\ensuremath{\rm ~\mus}\xspace}

\def\ps   {\ensuremath{\rm ~ps}\xspace}

\def\mus        {\ensuremath{~\mu{\rm s}}\xspace}    
\def\ps         {\ensuremath{{\rm ~ps}}\xspace}  

\def\to                 {\ensuremath{\rightarrow}\xspace}

\def\gsim{{~\raise.15em\hbox{$>$}\kern-.85em
          \lower.35em\hbox{$\sim$}~}\xspace}
\def\lsim{{~\raise.15em\hbox{$<$}\kern-.85em
          \lower.35em\hbox{$\sim$}~}\xspace}

\def \cp {{\it CP}\xspace}

\newcommand{\Bkspzg}{\ensuremath{\Bz \to \KS \piz \g}\xspace}
\newcommand{\Bksppg}{\ensuremath{\Bp \to \KS \pip \g}\xspace}

\def\Kstz  {\ensuremath{K^{*0}}\xspace}
\renewcommand{\Kstarz}{\ensuremath{K^{*}(892)^{0}}\xspace}

\newcommand{\dt}{\ensuremath{\Delta t}\xspace}

\newcommand{\de}{\ensuremath{\Delta E}\xspace}
\newcommand{\dw}{\ensuremath{\Delta w}\xspace}
\newcommand{\dmd}{\ensuremath{\Delta m_d}\xspace}
\newcommand{\mbc}{\ensuremath{M_{\text{bc}}}\xspace}

\newcommand{\cosTB}{\ensuremath{\cos\theta_B^*}\xspace}

\newcommand{\sPlot}{\ensuremath{_s\mathcal{P}lot}\xspace}

\def\Bztag {\ensuremath{\Bz_{\rm tag}}\xspace}

\def\Bzbtag {\ensuremath{\Bzb_{\rm tag}}\xspace}

\def \Ebeam  {\ensuremath{E^*_{\rm beam}}\xspace}
\def \mks    {\ensuremath{m(\pipi)}\xspace}

\def \Rsig {\ensuremath{R_{\rm sig}}\xspace}

\def \dw {\ensuremath{\Delta w}\xspace}

\def \sPlot {\ensuremath{_{s}\mathcal{P}lot}\xspace}
\def \scp {\ensuremath{S}\xspace}
\def \acp {\ensuremath{A}\xspace}
\def \ccp {\ensuremath{C}\xspace}

\def \mkpi {\ensuremath{m(\KS\piz)}\xspace}

\newboolean{wordcount}
\setboolean{wordcount}{false} 

\ifthenelse{\boolean{wordcount}}%
{\usepackage{bibentry} 
 \usepackage{comment} 
 \excludecomment{align*} 
 \excludecomment{align} 
 \excludecomment{equation*} 
 \excludecomment{equation} 
 \excludecomment{eqnarray*} 
 \excludecomment{eqnarray} 
 \excludecomment{acknowledgments} 
 \def\maketitle{} 
}{}

\begin{document}


\title{Measurement of \textit{CP} asymmetries in {\boldmath $\Bz\to\KS\piz\g$} decays at Belle II}

\ifthenelse{\boolean{wordcount}}{}{
  \author{I.~Adachi\,\orcidlink{0000-0003-2287-0173}} 
  \author{L.~Aggarwal\,\orcidlink{0000-0002-0909-7537}} 
  \author{H.~Ahmed\,\orcidlink{0000-0003-3976-7498}} 
  \author{H.~Aihara\,\orcidlink{0000-0002-1907-5964}} 
  \author{N.~Akopov\,\orcidlink{0000-0002-4425-2096}} 
  \author{A.~Aloisio\,\orcidlink{0000-0002-3883-6693}} 
  \author{N.~Anh~Ky\,\orcidlink{0000-0003-0471-197X}} 
  \author{D.~M.~Asner\,\orcidlink{0000-0002-1586-5790}} 
  \author{H.~Atmacan\,\orcidlink{0000-0003-2435-501X}} 
  \author{T.~Aushev\,\orcidlink{0000-0002-6347-7055}} 
% ...
\collaboration{The Belle II Collaboration}
}

\begin{abstract}
We report measurements of time-dependent \cp asymmetries in $\Bz\to\KS\piz\g$ decays based on a data sample of $(388\pm6)\times10^6$ \BB events collected at the \FourS resonance with the \belletwo detector.
The \belletwo experiment operates at the SuperKEKB asymmetric-energy \epem collider.
We measure decay-time distributions to determine \cp-violating parameters \scp and \ccp.
We determine these parameters for two ranges of $\KS\piz$ invariant mass: 
$\mkpi\in (0.8, 1.0)\gevcc$, which is dominated by $\Bz\to\Kstz(\to\KS\piz)\g$ decays,
and a complementary region $\mkpi\in (0.6, 0.8)\cup(1.0, 1.8)\gevcc$.
Our results have improved precision as compared to previous measurements and are consistent with theory predictions.

\end{abstract}

\maketitle

Flavor-changing neutral current decays of elementary particles are of great interest since they predominantly occur through quantum loop-level processes,
making them highly sensitive to phenomena beyond the Standard Model (SM).
Due to the heavy $b$ quark mass and enhanced loop contribution of the $t$ quark, processes involving the $b \to s$ quark transition are especially sensitive to physics beyond the SM (BSM).
Decays via the $b \to s \g$ radiative transition are similary important and have thus been studied both theoretically and experimentally~\cite{Bertolini:1990if,Cho:1993zb,Fujikawa:1993zu}.
In particular, the polarization of the final-state photon adds 
unique sensitivity to BSM physics~\cite{Atwood:1997zr,Atwood:2004jj,Blanke:2012tv,Becirevic:2012dx,Shimizu:2012zw,Kou:2013gna,Malm:2015oda,Eberl:2021ulg}.
In the SM, because $W$ bosons interact only with left-handed fermions, the $s$ quark from $b \to s \g$ (a $\Bzb$ decay) is left-handed and thus the outgoing photon must have negative helicity to conserve angular momentum along the decay axis.
Similarly, the outgoing photon from a $\Bz$ decay must have positive helicity.
The ``wrong'' photon polarization is possible only if the $s$ quark flips its chirality,
which suppresses this process by a factor of~$m_s/m_b$.

In coherent $B$-pair production via $\epem\to\FourS\to\BzBzb$,
the time-dependent decay rate of one $B$ meson, denoted $B_{\rm sig}$, decaying into $\KS(\to\pipi)\piz(\to\g\g)\g$, and the accompanying  $B$ meson, denoted $B_{\rm tag}$, decaying with flavor $q$ ($q=1$ for $\Bztag$ and $-1$ for $\Bzbtag$), is given by~\cite{Carter:1980hr,Carter:1980tk,Bigi:1981qs,FN:MixingCPV}
\begin{eqnarray}
\label{eq:model}
    \mathcal{P}(\dt,q) &=& \frac{e^{-|\dt|/\tau_{\Bz}}}{4\tau_{\Bz}}
    \{ 1+q\cdot [\scp\sin(\dmd\dt) \nonumber \\
    & &-\ccp\cos(\dmd\dt) ] \},
\end{eqnarray}
where $\dt \equiv t_{\rm sig} - t_{\rm tag}$ is the difference between the proper decay times of $B_{\rm sig}$ and $B_{\rm tag}$, 
\scp and \ccp are parameters characterizing mixing-induced and direct {\it CP} violation~\cite{footnote2}, respectively,
$\tau_{\Bz}$ is the $\Bz$ lifetime,
and $\Delta m_d$ is the mass difference between the two neutral $B$-meson mass eigenstates.
In the SM, the \scp value
is expected to be very small, as the different photon polarizations distinguish between otherwise identical final states originating from $\Bz$ or $\Bzb$ and thus preclude interference~\cite{Atwood:1997zr,Atwood:2004jj}.
However, if right-handed currents from BSM physics contribute, 
a $\Bz$ ($\Bzb$) could more easily emit negative (positive) helicity photons,
leading to sizable interference and 
$\scp\sim\mathcal{O}$(0.1)~\cite{Blanke:2012tv,Becirevic:2012dx,Shimizu:2012zw,Kou:2013gna,Malm:2015oda,Eberl:2021ulg}.
For resonant $\Bz\to\Kstarz(\to\KS\piz)\g$~\cite{FootnoteCC},
two SM calculations give $\scp=(-3.5\pm1.7)\times10^{-2}$~\cite{Matsumori:2005ax} and $(-2.3\pm1.6)\times10^{-2}$~\cite{Ball:2006eu}.
Nonresonant $\Bz\to\KS\piz\g$ decays could include a long-distance contribution from $b\to (c \cbar) s$ rescattering and should be measured separately~\cite{Grinstein:2004uu,Grinstein:2005nu}.
The \ccp value in these decays has not been reliably estimated~\cite{footnote2}.

Previously, 
the Belle and {\it BABAR\/} experiments
measured the \scp values in $\Bz\to\KS\piz\g$ decays with a precision of about 
30\%~\cite{Belle:2006pxp,BaBar:2008okc}.
The LHCb experiment also measured the photon polarization in $b\to s\g$ transitions~\cite{LHCb:2019vks,LHCb:2020dof,LHCb:2021byf}.
In addition, these experiments measured the direct {\it CP} asymmetry with 2\% precision~\cite{Belle:2017hum,BaBar:2009byi,LHCb:2012quo}.
These results are consistent with SM predictions, although BSM contributions cannot be excluded.
In this Letter, we report new measurements of time-dependent {\it CP} asymmetries in $\Bz\to\Kstarz\g$ and nonresonant $\Bz\to\KS\piz\g$ decays using 365\invfb of data~\cite{Belle-II:2024vuc}, corresponding to $(388\pm6)\times10^6$ $\BB$ events, recorded by the Belle~II experiment from 2019 to 2022~\cite{Belle-II:2010dht}.

Belle~II~\cite{Belle-II:2010dht} operates at the SuperKEKB collider~\cite{Akai:2018mbz}, which collides 7.0\gev electrons with 4.0\gev positrons.
The detector components most relevant for this measurement are a two-layer silicon-pixel detector~(PXD), a four-layer double-sided silicon-strip detector~(SVD)~\cite{Belle-IISVD:2022upf}, and a 56-layer central drift chamber~(CDC).
These detectors reconstruct tracks of charged particles and measure displaced vertices.
Only one sixth of the second PXD layer
was installed for the data analyzed here.
The symmetry axis of these cylindrical detectors, defined as the $z$ axis, is almost collinear with the electron beam direction.
The detector coverage is divided into three regions depending on the polar angle $\theta$:
the barrel for $32.2^\circ<\theta<128.7^\circ$, and the forward and backward endcaps for $12.4^\circ<\theta<31.4^\circ$ and $130.7^\circ<\theta<155.1^\circ$, respectively.
Surrounding the CDC
is a time-of-propagation counter~\cite{Kotchetkov:2018qzw} in the barrel and an aerogel-based ring-imaging Cherenkov counter in the forward endcap.
These detectors provide charged-particle identification (PID).
Surrounding the PID detectors is an electromagnetic calorimeter~(ECL) based on CsI(Tl) crystals that provides energy and timing measurements for photons and electrons.
The subdetectors described above are enclosed within
a superconducting solenoid that provides a 1.5~T magnetic field
oriented in the $z$ direction.

We use Monte Carlo simulation to
optimize event selection criteria, calculate reconstruction 
efficiencies, and study sources of background.
We generate
simulated signal samples of
$\Bz\to\Kstarz\g$, $\Bz\to K_2^{*0}(1430)\gamma$, $\Bz\to K^{*0}(1680)\gamma$,
and nonresonant $\Bz\to\KS\piz\g$~\cite{Kagan:1998ym}.
To model backgrounds, we generate samples of $\epem\to\BzBzb$, $\BpBm$, $\qqbar$ ($q=u,d,s,c$), and $\tautau$. 
We use \textsc{EvtGen}~\cite{Lange:2001uf} for hadronic decays, \textsc{KKMC}~\cite{kkmc} followed by fragmentation by \textsc{Pythia}~\cite{pythia} for $\qqbar$, and \textsc{Tauola}~\cite{Jadach:1990mz} for $\tau$ decays.
The detector response is simulated with \textsc{Geant4}~\cite{GEANT4:2002zbu}.
We analyze data and simulated events using the Belle~II software~\cite{basf2}.
We optimize~\cite{differential_evolution} selection criteria for mass windows, momentum, and boosted decision trees (BDTs), maximizing
$N^{}_s/\sqrt{N^{}_s+N^{}_b}$, where $N^{}_s$ and $N^{}_b$ are signal and background yields in simulation.

To reconstruct the prompt photon, we select a cluster of ECL hits with no associated track.
We require that the photon's energy in the center-of-mass (c.m.) frame is the highest in the event and exceeds $1.6\gev$.
The dominant background is from photons arising from $\piz$ and $\eta$ decays; 
these are rejected with a BDT-based algorithm~\cite{Belle-II:2024tru}.
We apply a selection on the BDT output that,
according to simulation,
rejects 83\% (45\%) of photons from $\piz$ ($\eta$) decays
while retaining 92\% (99\%) of prompt signal photons.

We reconstruct $\KS\to\pipi$ candidates using two oppositely charged tracks. 
The $\KS$ candidates are selected with an invariant mass in the range $|\mks-m_{\KS}|<34\mevcc$, where $m_{\KS}$ is the known $\KS$ mass~\cite{PDG}.
This range corresponds to $\pm2.6\sigma$ in resolution.
To further reduce background, we employ another BDT to form two classifiers: $\KS$-likeness and $\Lambda$-likeness, which are based on kinematic and PID information.
We apply selection criteria on these classifiers that retain 97\% of signal $\KS$ decays, 
according to simulation,
while rejecting 67\% of candidates due to $\Lambda$'s and 98\% of other background candidates.

We reconstruct $\piz\to\g\g$ candidates using pairs of ECL 
clusters with no associated tracks.
We require a cluster energy greater than 22.5 (20.0)\mev in the forward endcap (barrel and backward endcap).
Photon pairs having an invariant mass 
$m(\gamma\gamma)\in (104, 164)\mevcc$,
corresponding to $2\sigma$ in resolution, are selected as $\piz$ candidates.
We also require 
that the $\piz$ momentum exceeds
$430\mevc$.
This requirement retains 92\% of signal $\piz$'s,
according to simulation,
while rejecting 74\% of background~$\piz$'s.

We reconstruct $B_{\rm sig}$ candidates by combining $\g$, $\KS$, and $\piz$ candidates.
We select $B_{\rm sig}$ candidates using the beam-energy-constrained mass $\mbc\equiv\sqrt{E^{*2}_{\rm beam}/c^4-\vectorbold{p}^{*2}_B/c^2}$, and energy difference $\de\equiv E^*_{B}-\Ebeam$, where $\Ebeam$ is the beam energy and $E^*_B$ and $\vectorbold{p}^*_B$ are the reconstructed energy and momentum of the $B_{\rm sig}$ candidate.
All quantities are evaluated in the c.m.\ frame.
We retain $B_{\rm sig}$ candidates satisfying 
$5.20 <\mbc<5.29\gevcc$ and $-0.5 <\de<0.5\gev$.
Almost half (45\%) of events have multiple $B_{\rm sig}$ candidates, with 96\% of these due to multiple $\piz$ candidates.
To identify the correct $B_{\rm sig}$ candidate, 
we use a BDT-based classifier based on 
the properties of
the $\piz$ candidate~\cite{piz}.\nocite{55109}
We retain the $B_{\rm sig}$ candidate with the highest $\piz$ BDT classifier value.
If multiple candidates share the same $\piz$, we choose the candidate with the highest value of $\KS$-likeness.
These criteria
select the correct $B_{\rm sig}$ candidate in 85\% of 
simulated
events with multiple candidates.
The invariant mass of the $\KS\piz$ system, $\mkpi$, is required to be in the range 0.6--1.8\gevcc.
We define mass region~1~(MR1) as $\mkpi\in (0.8, 1.0)\gevcc$,
which is dominated by the $\Kstarz$ meson, and mass region~2~(MR2) 
as $\mkpi\in (0.6, 0.8)\cup(1.0, 1.8)\gevcc$. 

We measure the decay vertex positions of $B_{\rm sig}$ and $B_{\rm tag}$ in 
kinematic fits.
The $B_{\rm sig}$ vertex is determined by a fit to the entire decay chain~\cite{Belle-IIanalysissoftwareGroup:2019dlq}.
This fit includes the constraint that the $B_{\rm sig}$ trajectory be consistent with originating from the $\epem$ interaction point (IP).
The IP is measured regularly by averaging over $\epem\to\mumu$ events.
For the $B_{\rm tag}$ vertex~\cite{Dey:2020dsr}, the tracks used
must have a distance of closest approach to the IP within 0.5\cm 
in the $r$--$\phi$ plane and within 2.0\cm along the $z$ axis.
We also require that each track have at least one hit in each of the PXD, SVD, and CDC subdetectors, and a momentum greater than $50\mevc$.

We calculate the decay time
$\dt$ as $(\ell_{\rm sig}-\ell_{\rm tag})/\beta\gamma c$, where 
$\ell_{\rm sig}$ ($\ell_{\rm tag}$) is the $B_{\rm sig}$ ($B_{\rm tag}$) decay vertex
position projected along the $\FourS$ boost direction, and $\beta\gamma$ is the Lorentz boost factor of the $\FourS$ in the lab frame.
This calculation applies to events satisfying the following ``TD'' (time-dependent) criteria.
For the $B_{\rm sig}$ vertex,
both pions from the $\KS$ must have at least one SVD hit; 
the $\chi^2$ of the vertex fit must be 
less than 30 (100) for tracks (overall);
and the uncertainty on $\ell_{\rm sig}$,
$\sigma_{\ell_{\rm sig}}$, must be less than $500\mum$.
For the $B_{\rm tag}$ vertex, the reduced $\chi^2$ of the vertex fit 
must be less than 100, and $\sigma_{\ell_{\rm tag}}$ must be
less than $500\mum$.
Events that do not satisfy these criteria are classified as ``TI'' (time-integrated);
they are included in the fit (described below) to improve the precision on \ccp.

The flavor {\it q} of $B_{\rm tag}$ is determined using a combination of BDT algorithms~\cite{Belle-II:flavortag}.
The combination outputs the product $q\cdot r$, where $r$ is a quality factor that ranges from zero for no flavor information to 1.0 for unambiguous flavor assignment.
As the tagging efficiency and signal purity depend on $r$, we divide the data into seven $r$ bins, each containing a similar number of events. 
For each $r$ bin, the wrong-tag fraction $w$ and the difference $\dw$ between $\Bz$ and $\Bzb$ tags are determined using flavor-specific $B$ decays~\cite{Belle-II:2023nmj}.
The effective tagging efficiency, $\varepsilon_{\rm eff} = \Sigma_i \varepsilon_{\rm tag}^i(1-2w_i)^2$, is $(31.69\pm0.35)\%$,
where $\varepsilon_{\rm tag}^i$ is the probability for a $\Bz\to\KS\piz\g$ decay to be flavor-tagged with $r$ in the $i$th bin. 

We train two BDT classifiers to discriminate signal 
from $\qqbar$ background,
separately for MR1 and MR2 candidates. The classifiers
use event topology variables~\cite{qqbar},
\nocite{FoxWolfram,Bevan:2014iga,BRANDT196457,PhysRevLett.39.1587,KSFW,cleocones}
and the classifier thresholds are optimized separately 
for each $r$ bin. According to simulation, these thresholds retain 
77\% (74\%) of signal decays while rejecting 95\% (89\%) of 
$\qqbar$ background for MR1 (MR2).

As a control mode, we reconstruct the decay \Bksppg,
which does not exhibit mixing-induced {\it CP} violation.
This decay, when the primary \pip track is ignored, has the same vertex resolution as the signal \Bkspzg decay, as the contribution of the \piz candidate to the signal vertex resolution is negligible.
 
\begin{figure}[htb]
\centering
\includegraphics[width=0.49\textwidth]{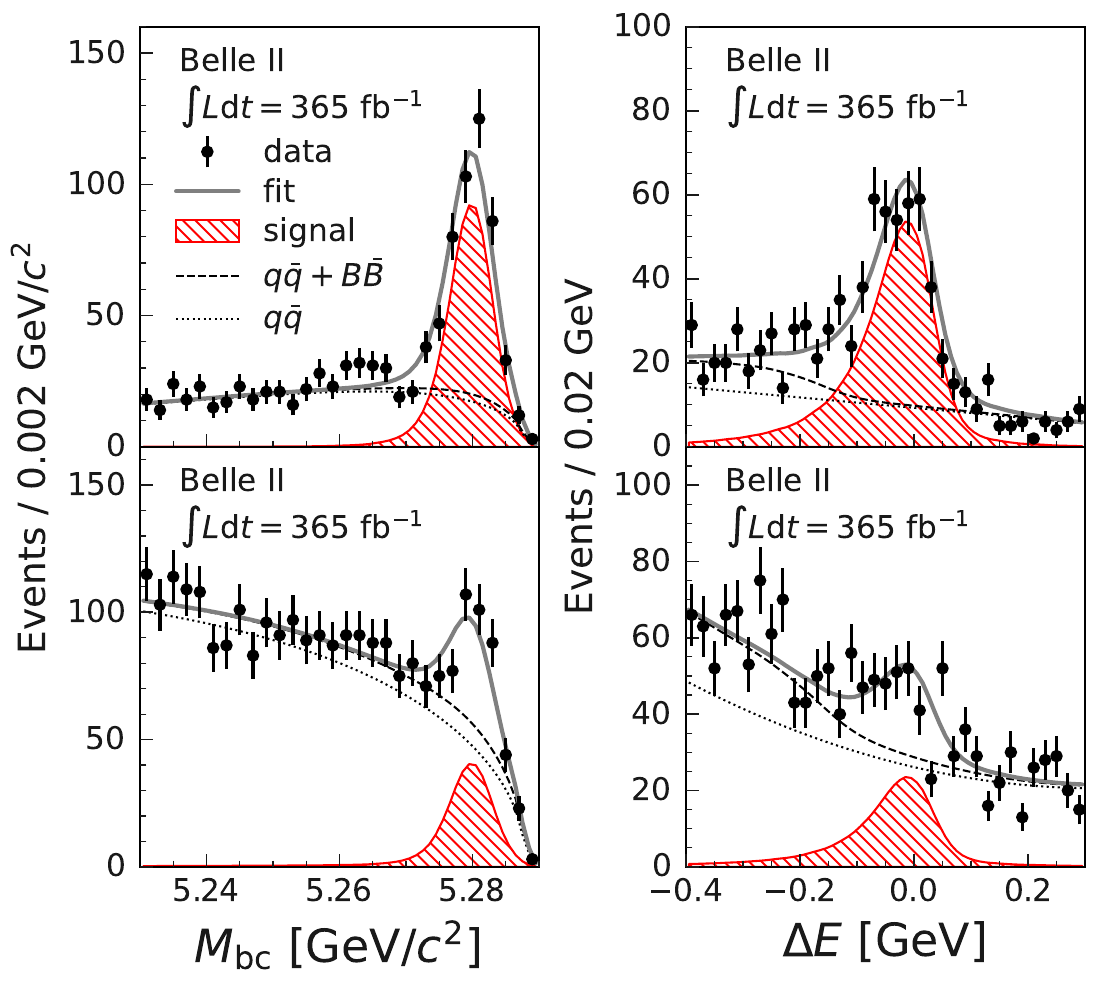}
  \caption{
  Distributions of \mbc (left) and \de (right) for 
  MR1 (top) and MR2 (bottom), with fit results overlaid. The \mbc (\de)
  distribution corresponds to the \de (\mbc) signal-enhanced region.}
  \label{fig_SEpreLS1}
\end{figure}

The signal and background yields are determined from an unbinned maximum-likelihood fit to the $\mbc$--$\de$ distribution.
We fit events in the ranges $5.23<\mbc<5.29\gevcc$ and $-0.4<\de<0.3\gev$.
We combine the TD and TI samples, as they (both for data and for simulated signal and background events) show negligible differences in these variables.
We model the $\mbc$--$\de$ probability density functions (PDFs) for signal and $\BB$ background using simulation.
Kernel density estimation (KDE)~\cite{Cranmer:2000du} is used to account for $\mbc$--$\de$ correlations.
The $\qqbar$ background is modeled with
the product of an ARGUS function~\cite{ARGUS:1990hfq} for $\mbc$ 
and a second-order polynomial for $\de$.
The signal and $\BB$ yields, and the $\qqbar$ shape parameters, are 
floated in the fit, while the total yield is fixed.
To calculate signal yields, we define a signal-enhanced
region $5.27 <\mbc<5.29\gevcc$ and $-0.2 <\de<0.1\gev$.
For $\Bz\to\Kstarz\g$ events, the efficiency in this region is $(22.8\pm 0.1)$\% for MR1.
The fitted $\mbc$ and $\de$ distributions and 
projections of the fit result are shown in \cref{fig_SEpreLS1}.
The resulting signal and $B\overline{B}$ yields are listed in~\cref{tab:yield}.

\begin{table}[htb]
    \centering
    \caption{
    Signal and control sample yields in the signal-enhanced region.
    Also listed is the signal-to-background ratio
    ($N^{}_s/N^{}_b$), where $N^{}_b$
    includes all backgrounds. }
    \begin{tabular}{l c c c}
    \hline\hline
        Sample & Signal yield & \BB bkg yield & 
        $N^{}_s/N^{}_b$ 
        \\ \hline
        $\Bz\to\KS\piz\g$ in MR1     & $385\pm24$ & $20\pm\phantom{1}8$ & 2.36\\
        $\Bz\to\KS\piz\g$ in MR2 & $171\pm23$ &   $69\pm19$  & 0.34\\
        \Bksppg & $843\pm34$ &   $55\pm10$  & 2.68\\
        \hline\hline
    \end{tabular}
    \label{tab:yield}
\end{table}

The $\dt$ PDF consists of signal and background components weighted by their fractions as determined from the $\mbc$--$\de$ fit.
For the TD category, the PDF is described as 
\begin{linenomath}
\begin{align}
    P_{\rm TD}(\dt,q) &={} \int^\infty_{-\infty}\Bigl[f_{\rm sig}\mathcal{P}_{\rm sig}(\dt',q)\Rsig(\dt-\dt')\nonumber\\
    +(1-&f_{\rm sig})f_{\BB}\mathcal{P}_{\BB}(\dt',q)R_{\BB}(\dt-\dt')\Bigr]d\dt'\nonumber\\
    &+(1-f_{\rm sig})(1-f_{\BB})
    \mathcal{P}_{\qqbar}(\dt)\,,\\
    \label{eq:TDfit}
    \mathcal{P}_{\rm sig}(\dt',q) &= \frac{1}{4\tau_{\Bz}}e^{-\frac{|\dt'|}{\tau_{\Bz}}}\Bigl\{1-q\dw+q(1-2w)\nonumber\\
    \times\bigl[\scp&\sin(\dmd\dt')-\ccp\cos(\dmd\dt')\bigr]\Bigr\},\\
    \mathcal{P}_{\BB}(\dt',q) &= \frac{1}{4\tau_{\BB}}e^{-\frac{|\dt'|}{\tau_{\BB}}}(1-q\dw) \,, 
\end{align}
\end{linenomath}
where $f_{\rm sig}$ is the event-by-event signal probability, $f_{\BB}$ is the $\BB$ background probability relative to that of all background components, $\mathcal{P}_{\rm sig}(\dt',q)$ and $\mathcal{P}_{\BB}(\dt',q)$ are the signal and $\BB$ background PDFs taking into account the effect of $w$ and $\dw$,  $\mathcal{P}_{\qqbar}(\dt)$ is the $\qqbar$ PDF, and $\Rsig$ and $R_{\BB}$ are the proper-time resolution functions for signal and $\BB$ background~\cite{resolution}.
We fix $\tau_{\Bz}$ and $\Delta m_d$ to their known values~\cite{PDG} and the effective lifetime of the $\BB$ background, $\tau_{\BB}$, to the value determined from simulation.
The fractions $f_{\rm sig}$ and $f_{\BB}$
are taken from the previous fit.
The only floated parameters are $S$ and $C$.
 
The resolution functions, $\Rsig$ and $R_{\BB}$, are described by convolving three components:
detector resolution for $B_{\rm sig}$ and $B_{\rm tag}$ decay vertices; 
bias due to the secondary decay vertex of intermediate charm states in $B_{\rm tag}$ decays;
and a correction to the boost factors due to the small momenta of the $B$ mesons in the c.m.\ frame.
The last component depends on $\cosTB$, where 
$\theta_B^*$ is the angle in the c.m.\ frame between the $\Bz$ momentum 
and the $\FourS$ boost direction~\cite{Tajima:2003bu}.
We model the detector resolution for $B_{\rm sig}$ ($B_{\rm tag}$) vertex on an event-by-event basis,
including the dependence on $\sigma_{\ell_{\rm sig}}$ ($\sigma_{\ell_{\rm tag}}$) and the $\chi^2$
from the
vertex fit.
The resolution function parameters are determined from simulated samples generated with five parameters of a helix-like trajectory~\cite{BelleIITrackingGroup:2020hpx} calibrated from data.
The data used consist of cosmic ray events in which a cosmic ray track traversing the PXD or SVD is reconstructed as two outgoing tracks from the IP. 

For $\BB$ background, the $B_{\rm sig}$ vertex resolution is the same as that of the signal component due to the high purity of $\KS$ candidates, while the $B_{\rm tag}$ vertex suffers from contamination by tracks coming from the background $B_{\rm sig}$ decay.
We adjust $\tau_{\BB}$ in $\mathcal{P}_{\BB}$ and parameters of the 
$\BB$ resolution function 
to account for the smeared vertex position and the typically shorter decay time difference.
We model $\mathcal{P}_{\qqbar}(\dt)$ using three Gaussian functions, whose shape parameters are determined from a fit to an $\mbc$--$\de$ sideband in data defined as $5.23 <\mbc<5.255\gevcc$, $-0.5 <\de<0.3\gev$, and $(\mbc - 5.23 \gevcc) < (\de + 0.45\gev)/14c^2$.

The PDF $P_{\rm TD}$ depends on $\cosTB$, $\sigma_{\ell_{\rm sig}}$, $\sigma_{\ell_{\rm tag}}$, the vertex-fit $\chi^2$'s, and $r$ (through $\dw$)~\cite{resolution}.
If signal and background events are distributed differently in these variables, then including them in the PDF can introduce bias~\cite{Punzi}.
Among these variables, only $r$ and $\cosTB$ differ noticeably between signal and backgrounds, and to alleviate bias we include additional PDFs for them in the likelihood function~\cite{Punzi}.
The $r$ and $\cosTB$ distributions for signal and \BB background are determined from simulation, while those for $\qqbar$ background are taken from the $\mbc$--$\de$ sideband in data.

For the TI category, the PDF is expressed as
\begin{linenomath}
\begin{align}
    P_{\rm TI}(q) ={}& f_{\rm sig}\left(\frac{1}{2}\right)
    \qty[1-q\dw-q(1-2w)\frac{\ccp}{1+\dmd^2\tau_{\Bz}^2}]\nonumber\\
    &+(1-f_{\rm sig})f_{\BB}\left(\frac{1-q\dw}{2}\right)\nonumber\\
    &+(1-f_{\rm sig})(1-f_{\BB})\left(\frac{1}{2}\right) ,
\label{eq:TIfit}
\end{align}
\end{linenomath}
where only the $C$ parameter is extracted in the fit.

\begin{figure}[tb]
\centering
\includegraphics[width=0.49\textwidth]{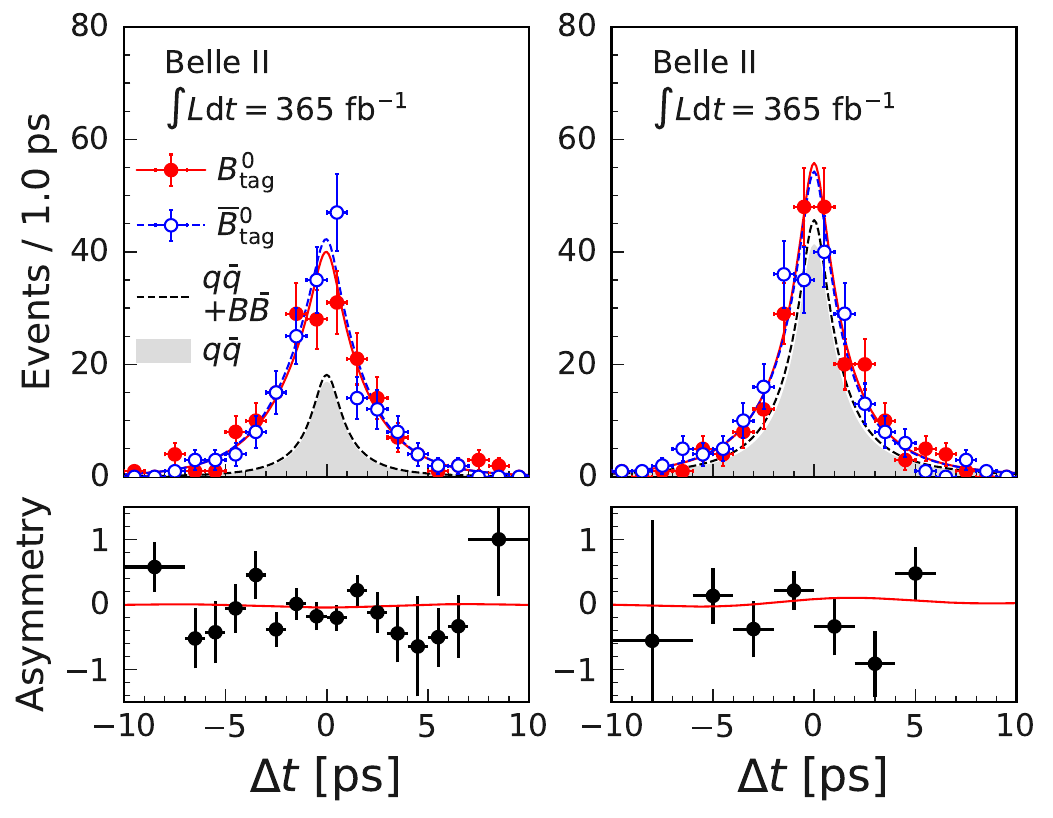}
\caption{
\dt distributions for MR1 (left) and MR2 (right)
with the fit result superimposed. The asymmetry $[N^{}_s(\Bztag)-N^{}_s(\Bzbtag)]/[N^{}_s(\Bztag)+N^{}_s(\Bzbtag)]$ is calculated in each \dt bin using \sPlot~\cite{Pivk:2004ty} and plotted in the bottom panels along with
the fit result.
}
\label{fig_cp_control_preLS1}
\end{figure}

We simultaneously fit the TD and TI samples in the signal-enhanced region, floating the common $C$ parameter.
The results are 
$\scp=0.00\,^{+0.27}_{-0.26}$ and $\ccp=0.10\pm 0.13$ for MR1 events,
and $\scp=0.04\,^{+0.45}_{-0.44}$ and $\ccp=-0.06\pm 0.25$ for MR2 events. The 
statistical correlation between $S$ and $C$ is $-0.005$ in MR1 and $+0.011$ in MR2.
\Cref{fig_cp_control_preLS1} shows the $\dt$ distribution 
along with the fit result.
No significant time-dependent asymmetries are observed.

We perform various cross-checks to confirm the validity of our fit procedure.
We fit for the $\Bz$ lifetime in the MR1 and MR2 samples and obtain $1.55 \pm 0.14\ps$ and $1.58 \pm 0.24\ps$, respectively, which are consistent with the world average~\cite{PDG}.
We also perform studies of the control mode $\Bksppg$.
The lifetime fit obtains $1.68 \pm 0.09\ps$, which is consistent with the $\Bp$ lifetime.
The difference between $\Bp$ vertices with and without the $\pip$ track is consistent with the $B_{\rm sig}$ vertex resolution function.
We fit for {\it CP} violation and obtain $\scp = 0.05 \pm 0.09$ and $\ccp = 0.03 \pm 0.05$, which are consistent with $\scp=0$ and the world average $\ccp=0.014\pm0.018$~\cite{PDG}.

We consider various sources of systematic uncertainty.
We repeat the analysis with shifted scaling factors for the track momentum and cluster energy and take the deviation from the nominal results as the uncertainty.
In a similar manner, we treat four sources of uncertainties for the vertex measurements:
possible detector misalignments, 
imperfect understanding of the IP profile 
measurement~\cite{Belle-II:2023nmj}, 
corrections to uncertainties on the track helix parameters,
and the TD criteria.
We assess the effect of uncertainties in $w$ and $\dw$ by varying them by their uncertainties and refitting the data.
We also simulate the effect of a difference in $\varepsilon_{\rm tag}^i$ between $\Bztag$ and $\Bzbtag$ decays.
The uncertainties due to fixed parameters for $\mbc$, $\de$, and $\cosTB$ PDFs and $r$-bin fractions in signal modeling, including $f_{\rm sig}$ and $f_{\BB}$, are also evaluated by refitting.
We simulate potential mismodeling of $\mbc$--$\de$ distributions and potential bias from the modeling of $\sigma_{\ell_{\rm sig}}$, $\sigma_{\ell_{\rm tag}}$ and the vertex-fit $\chi^2$'s~\cite{Punzi}.
The uncertainties due to limited statistics of the dataset used for KDE are evaluated with the bootstrap method~\cite{bootstrap}.

We consider uncertainties from the modeling and parameters of the $\dt$ resolution function by varying these parameters by their uncertainties and refitting.
The differences between these results and our nominal result are taken as uncertainties.
We evaluate the uncertainties arising from $\tau_{\Bz}$ and $\Delta m_d$ in a similar manner~\cite{PDG}.

We study the impact of an asymmetry in $\BB$ backgrounds.
We generate samples with asymmetric backgrounds and refit using our nominal (symmetric) $\BB$ background PDF.
The change in the fit results is assigned as an uncertainty.
The asymmetries simulated correspond to the world average values of $S$, $C$ for $b \to s\qqbar$ decays, and to maximum values of $S$, $C$ ($\pm1$) for $b \to s \g$ decays.

We evaluate the bias from tag-side interference~\cite{Long:2003wq} assuming $\scp=\ccp=0$.
The dominant uncertainties come from the scaling factors for cluster energy, the vertex quality selection criteria, the $\BB$ background asymmetry, and the tag-side interference.
The total systematic uncertainties, listed in \cref{tab:systematics}, are calculated as the sum in quadrature of all individual systematic uncertainties.

\begin{table}[htb]
\centering
  \caption{Summary of systematic uncertainties.}
  \begin{tabular}{l c c c c}
  \hline\hline
   & \multicolumn{2}{c}{$\Kstz\g$} & \multicolumn{2}{c}{$\KS\piz\g$} \\
    Source & $\scp$ & $\ccp$ & $\scp$ & $\ccp$ \\ \hline
    $E$ and $p$ scales         & $\pm0.017$           & $\pm0.015$           & $\pm0.083$           & $\pm0.047$ \\
    Vertex measurement         & $\pm0.021$           & $\pm0.009$           & $\pm0.023$           & $\pm0.036$ \\
    Flavor tagging             & $\pm0.005$           & $^{+0.012}_{-0.009}$ & $^{+0.008}_{-0.009}$ & $^{+0.013}_{-0.009}$ \\
    Signal modeling            & $\pm0.003$           & $\pm0.003$           & $\pm0.032$           & $\pm0.013$ \\
    \dt resolution function    & $\pm0.014$           & $\pm0.009$           & $\pm0.031$           & $\pm0.013$ \\
    $\tau_{\Bz}$ and \dmd      &   $<0.001$           &   $<0.001$           & $\pm0.003$           &   $<0.001$ \\
    \BB background asym.       & $^{+0.007}_{-0.008}$ & $\pm0.011$           & $^{+0.030}_{-0.026}$ & $^{+0.049}_{-0.051}$ \\
    Tag-side interference      &   $\pm0.003$           &   $\pm0.028$           &   $\pm0.003$           &   $\pm0.028$           \\ \hline
    Total                      & $\pm0.032$           & $^{+0.037}_{-0.038}$ & $^{+0.102}_{-0.101}$ & $\pm0.085$ \\
  \hline\hline
  \end{tabular}
\label{tab:systematics}
\end{table}

In summary, we measure the time-dependent {\it CP} asymmetry in $\Bz\to\KS\piz\g$ decays using $(388\pm6)\times10^6$ $\BB$ events collected by the Belle~II detector.
These decays are sensitive to right-handed currents arising from BSM physics.
We perform these measurements for two $\mkpi$ mass regions
corresponding to $\Kstz\to\KS\piz$ and 
nonresonant decays.
We obtain {\it CP}-violating parameters
\begin{linenomath}
\begin{equation}
\scp=0.00^{+0.27}_{-0.26}\pm0.03\,, \quad 
\ccp=0.10\pm0.13\pm0.04
\end{equation}
\end{linenomath}
for the $\Kstz$ resonant region and
\begin{linenomath}
\begin{equation}
\scp=0.04^{+0.45}_{-0.44}\pm0.10\,, \quad
\ccp=-0.06\pm0.25\pm0.09
\end{equation}
\end{linenomath}
for the nonresonant region, where the uncertainties are statistical and systematic, respectively.
The measured \scp values agree with SM predictions~\cite{Matsumori:2005ax,Ball:2006eu} within one standard deviation.
Our results have improved precision with respect to previous measurements~\cite{Belle:2006pxp,BaBar:2008okc}.
The improvements mostly result from a refined $\KS$ identification
algorithm and a large-acceptance silicon vertex detector~\cite{Belle-IISVD:2022upf}.
The improved precision should further constrain the BSM parameter space~\cite{Atwood:1997zr,Atwood:2004jj,Blanke:2012tv,Becirevic:2012dx,Shimizu:2012zw,Kou:2013gna,Malm:2015oda,Eberl:2021ulg}.

\begin{acknowledgments}
This work, based on data collected using the Belle II detector, which was built and commissioned prior to March 2019, was supported by
Higher Education and Science Committee of the Republic of Armenia Grant No.~23LCG-1C011;
Australian Research Council and Research Grants
No.~DP200101792, 
No.~DP210101900, 
No.~DP210102831, 
No.~DE220100462, 
No.~LE210100098, 
and
No.~LE230100085; 
Austrian Federal Ministry of Education, Science and Research,
Austrian Science Fund
No.~P~31361-N36
and
No.~J4625-N,
and
Horizon 2020 ERC Starting Grant No.~947006 ``InterLeptons'';
Natural Sciences and Engineering Research Council of Canada, Compute Canada and CANARIE;
National Key R\&D Program of China under Contract No.~2022YFA1601903,
National Natural Science Foundation of China and Research Grants
No.~11575017,
No.~11761141009,
No.~11705209,
No.~11975076,
No.~12135005,
No.~12150004,
No.~12161141008,
and
No.~12175041,
and Shandong Provincial Natural Science Foundation Project~ZR2022JQ02;
the Czech Science Foundation Grant No.~22-18469S;
European Research Council, Seventh Framework PIEF-GA-2013-622527,
Horizon 2020 ERC-Advanced Grants No.~267104 and No.~884719,
Horizon 2020 ERC-Consolidator Grant No.~819127,
Horizon 2020 Marie Sklodowska-Curie Grant Agreement No.~700525 ``NIOBE''
and
No.~101026516,
and
Horizon 2020 Marie Sklodowska-Curie RISE project JENNIFER2 Grant Agreement No.~822070 (European grants);
L'Institut National de Physique Nucl\'{e}aire et de Physique des Particules (IN2P3) du CNRS
and
L'Agence Nationale de la Recherche (ANR) under grant ANR-21-CE31-0009 (France);
BMBF, DFG, HGF, MPG, and AvH Foundation (Germany);
Department of Atomic Energy under Project Identification No.~RTI 4002,
Department of Science and Technology,
and
UPES SEED funding programs
No.~UPES/R\&D-SEED-INFRA/17052023/01 and
No.~UPES/R\&D-SOE/20062022/06 (India);
Israel Science Foundation Grant No.~2476/17,
U.S.-Israel Binational Science Foundation Grant No.~2016113, and
Israel Ministry of Science Grant No.~3-16543;
Istituto Nazionale di Fisica Nucleare and the Research Grants BELLE2;
Japan Society for the Promotion of Science, Grant-in-Aid for Scientific Research Grants
No.~16H03968,
No.~16H03993,
No.~16H06492,
No.~16K05323,
No.~17H01133,
No.~17H05405,
No.~18K03621,
No.~18H03710,
No.~18H05226,
No.~19H00682, 
No.~20H05850,
No.~20H05858,
No.~22H00144,
No.~22K14056,
No.~22K21347,
No.~23H05433,
No.~26220706,
and
No.~26400255,
the National Institute of Informatics, and Science Information NETwork 5 (SINET5), 
and
the Ministry of Education, Culture, Sports, Science, and Technology (MEXT) of Japan;  
National Research Foundation (NRF) of Korea Grants
No.~2016R1\-D1A1B\-02012900,
No.~2018R1\-A2B\-3003643,
No.~2018R1\-A6A1A\-06024970,
No.~2019R1\-I1A3A\-01058933,
No.~2021R1\-A6A1A\-03043957,
No.~2021R1\-F1A\-1060423,
No.~2021R1\-F1A\-1064008,
No.~2022R1\-A2C\-1003993,
and
No.~RS-2022-00197659,
Radiation Science Research Institute,
Foreign Large-Size Research Facility Application Supporting project,
the Global Science Experimental Data Hub Center of the Korea Institute of Science and Technology Information
and
KREONET/GLORIAD;
Universiti Malaya RU grant, Akademi Sains Malaysia, and Ministry of Education Malaysia;
Frontiers of Science Program Contracts
No.~FOINS-296,
No.~CB-221329,
No.~CB-236394,
No.~CB-254409,
and
No.~CB-180023, and SEP-CINVESTAV Research Grant No.~237 (Mexico);
the Polish Ministry of Science and Higher Education and the National Science Center;
the Ministry of Science and Higher Education of the Russian Federation
and
the HSE University Basic Research Program, Moscow;
University of Tabuk Research Grants
No.~S-0256-1438 and No.~S-0280-1439 (Saudi Arabia);
Slovenian Research Agency and Research Grants
No.~J1-9124
and
No.~P1-0135;
Agencia Estatal de Investigacion, Spain
Grant No.~RYC2020-029875-I
and
Generalitat Valenciana, Spain
Grant No.~CIDEGENT/2018/020;
National Science and Technology Council,
and
Ministry of Education (Taiwan);
Thailand Center of Excellence in Physics;
TUBITAK ULAKBIM (Turkey);
National Research Foundation of Ukraine, Project No.~2020.02/0257,
and
Ministry of Education and Science of Ukraine;
the U.S. National Science Foundation and Research Grants
No.~PHY-1913789 
and
No.~PHY-2111604, 
and the U.S. Department of Energy and Research Awards
No.~DE-AC06-76RLO1830, 
No.~DE-SC0007983, 
No.~DE-SC0009824, 
No.~DE-SC0009973, 
No.~DE-SC0010007, 
No.~DE-SC0010073, 
No.~DE-SC0010118, 
No.~DE-SC0010504, 
No.~DE-SC0011784, 
No.~DE-SC0012704, 
No.~DE-SC0019230, 
No.~DE-SC0021274, 
No.~DE-SC0021616, 
No.~DE-SC0022350, 
No.~DE-SC0023470; 
and
the Vietnam Academy of Science and Technology (VAST) under Grants
No.~NVCC.05.12/22-23
and
No.~DL0000.02/24-25.

These acknowledgements are not to be interpreted as an endorsement of any statement made
by any of our institutes, funding agencies, governments, or their representatives.

We thank the SuperKEKB team for delivering high-luminosity collisions;
the KEK cryogenics group for the efficient operation of the detector solenoid magnet;
the KEK computer group and the NII for on-site computing support and SINET6 network support;
and the raw-data centers at BNL, DESY, GridKa, IN2P3, INFN, and the University of Victoria for off-site computing support.

\end{acknowledgments}

\bibliographystyle{myapsrev4-2} 
\ifthenelse{\boolean{wordcount}}%
{ \nobibliography{references} }
{ \bibliography{references} }

\begin{thebibliography}{66}
\bibitem [{FN:()}]{FN:MixingCPV}%
  \BibitemOpen
  \href@noop {} {}\bibinfo {note} {Here, we neglect \cp violation in \Bz--\Bzb
  mixing.}\BibitemShut {Stop}%
\bibitem [{foo()}]{footnote2}%
  \BibitemOpen
  \href@noop {} {}\bibinfo {note} {The coefficients $(\scp,-\ccp)$ are written
  $(\scp, \acp)$ elsewhere.}\BibitemShut {Stop}%
\bibitem [{Foo()}]{FootnoteCC}%
  \BibitemOpen
  \href@noop {} {}\bibinfo {note} {Charge-conjugate modes are included unless
  explicitly stated otherwise.}\BibitemShut {Stop}%
\end{thebibliography}

\end{document}